\documentclass[conference]{IEEEtran}
\IEEEoverridecommandlockouts
\usepackage{cite}
\usepackage{amsmath,amssymb,amsfonts}
\usepackage{algorithmic}
\usepackage{graphicx}
\usepackage{textcomp}
\usepackage{xcolor}
\usepackage{siunitx} 
\usepackage{balance}
\usepackage{hyperref}
\usepackage{subfigure}
\usepackage{array}
\usepackage{booktabs}
\usepackage{multirow}
\usepackage{colortbl}
\usepackage{tabularx}
\definecolor{lightgray}{gray}{0.91}
\usepackage{setspace}
\def\BibTeX{{\rm B\kern-.05em{\sc i\kern-.025em b}\kern-.08em
		T\kern-.1667em\lower.7ex\hbox{E}\kern-.125emX}}
\usepackage{tikz}
\usetikzlibrary{arrows}
\tikzstyle{int}=[draw, fill=white!20, minimum width=3cm, minimum height=1cm]

\def\BibTeX{{\rm B\kern-.05em{\sc i\kern-.025em b}\kern-.08em
    T\kern-.1667em\lower.7ex\hbox{E}\kern-.125emX}}
\begin{document}

\title{General Molecular Communication Model in Multi-Layered Spherical Channels}
\author{\IEEEauthorblockN{Mitra Rezaei\IEEEauthorrefmark{1},
Michael Chappell\IEEEauthorrefmark{1},
Adam Noel\IEEEauthorrefmark{1}\IEEEauthorrefmark{2}, 
~\IEEEmembership{Senior Member,~IEEE}.}
\IEEEauthorblockA{\IEEEauthorrefmark{1}School of Engineering, University of Warwick, Coventry, UK.}
\IEEEauthorblockA{\IEEEauthorrefmark{2}Department of Electrical and Computer Engineering, Memorial University, St. John's, Canada.}
}

\maketitle

\begin{abstract}
Spherical multi-layered structures are prevalent in numerous biological systems and engineered applications, including tumor spheroids, layered tissues, and multi-shell nanoparticles for targeted drug delivery. Despite their widespread occurrence, there remains a gap in modeling particle propagation through these complex structures from a molecular communication (MC) perspective. This paper introduces a generalized analytical framework for modeling diffusion-based molecular communication in multi-layered spherical environments. The framework is capable of supporting an arbitrary number of layers and flexible transmitter-receiver positioning. As an example, the detailed formulation is presented for the three-layer sphere, which is particularly relevant for different biological models such as tumor spheroids. The analytical results are validated using particle-based simulation (PBS) in scenarios that have short inter-layer distances. The findings reveal that the characteristics of each layer significantly impact molecule propagation throughout the entire structure, making their consideration crucial for designing targeted therapies and optimizing drug delivery systems.
\end{abstract}


\section{Introduction}
\label{Sec:Introduction}
Diffusion-based molecular communication (MC), in which information is transmitted through Brownian motion of propagating molecules, has gained considerable attention in recent years \cite{8742793}. While most MC studies focus on simple, homogeneous channels \cite{8742793}, many biological environments comprise multiple layers with distinct diffusion properties. Examples include cellular microfluidic systems, respiratory membranes for oxygen and carbon dioxide diffusion, the blood-brain barrier in drug transport, and tumor micro-environments.

The majority of MC literature simplifies these complex, layered biological media into single homogeneous channels with constant viscosity and temperature, which results in a uniform diffusion coefficient \cite{mustam2017multilayer}. However, this approximation fails to capture the intricate dynamics of multi-layered systems, which can lead to inaccurate models and designs.
In recent years, a few attempts have been made to model MC systems within multi-layered geometries. 
Among the earliest MC studies to consider stratified layers, \cite{Composite} developed an MC channel model for multi-layered composite structures with \textit{parallel} layers of different diffusivity, where the outermost layers are considered semi-infinite. Further investigations on diffusion in parallel multi-layer structures can be found in \cite{carr2016semi,rodrigo2016solution,pontrelli2013local}.
However, the mathematical analysis of signal propagation in \textit{non-parallel} multi-layer geometries (e.g., layered cylinders and spheres) is still underexplored, even though such structures are common in nature. In the following, we review the relevant literature.

In \cite{jain2022theoretical}, the authors addressed the cylindrical geometry of arteries while investigating drug transport and retention in arterial walls following release from drug-coated balloons. Their analytical solution accounts for the coupled effects of diffusion, advection, and reaction processes. This approach contrasts earlier studies, such as \cite{pontrelli2013local}, which simplified the cylindrical artery to a one-dimensional parallel layered model.

The release of drugs from multi-layered coated capsules has been investigated in finite \cite{carr2018modelling} and semi-infinite \cite{kaoui2018mechanistic} media, with both studies considering imperfect boundary conditions that result in jump conditions at layer interfaces. The authors in \cite{carr2018modelling}, assuming uniform initial conditions, examined the effects of device geometry and coating resistance on the release mechanism. Extending this work, the authors in \cite{kaoui2018mechanistic} analyzed both desorbing and absorbing capsules by reversing the initial mass distribution and drug transport direction. They further improved the model by incorporating spatially-dependent initial conditions. Furthermore, they explored the solution's sensitivity to the coating mass transfer coefficient, which is a parameter dependent on the diffusivity and thickness of the coating layer, as an effective rate-controlling factor.

Recently, the authors of \cite{mahesh2022mathematical} developed a three-dimensional (3-D) numerical model to investigate interstitial fluid flow and nanoparticle distribution following intratumoral injection. Their model, which assumed perfect boundary conditions at interfaces, accounts for the multi-layer structure of spherical tumors. Their study examined several factors that influence nanoparticle distribution, such as particle size and the number of injection sites.

In \cite{al2020modelling}, the authors presented a numerical model for an MC system within spherical tumors, with a tumor implant as the transmitter, the tumor microenvironment as the channel, and malignant cells as the receiver. The study focused on a post-treatment tumor structure following partial ablation, a technique that destroys some cancer cells. This was modeled as a two-layer system: a central ablated zone of dead cells surrounded by a non-ablated zone within normal tissue. The model considered different diffusion coefficients for each layer while assuming perfect boundary conditions between interfaces. The authors studied the impact of varying release rates on molecular concentration both within and outside the tumor. Additionally, they examined toxicity levels in surrounding tissues under different extents of cell death. Their findings demonstrated that combining an implantable drug delivery system with thermal ablation in solid tumors results in high therapeutic efficacy with minimal toxicity to normal tissues.

In this study, we establish an innovative MC system model that addresses the complexity of multi-layered channels. We aim to develop a mathematical framework for particle propagation within multi-layered spherical structures, regardless of the number of layers or the positioning of transmitters and receivers. To the best of our knowledge, this is the first study to propose such a generalized model for diffusion problems in multi-layered spheres.
The versatility of our approach facilitates its application to various diffusion-dominated mass transfer problems. However, our focus in this paper is on MC systems within large spheroids -- 3-D cellular aggregations in a spherical shape -- which are fundamental components in organ-on-chip systems. Large spheroids are ideal examples due to their complex, multi-layered structure: an outer layer of loosely-attached cells surrounds intermediate layers with tighter cell packing and a denser extracellular matrix. This arrangement hinders oxygen diffusion to the core, potentially causing necrosis (death) of the cells in the center \cite{nath2016three}. We provide a detailed mathematical presentation of a three-layer model, given its prevalence in applications such as tumor modeling and drug delivery systems.
To validate our analytical results, we use particle-based simulation (PBS). Our PBS approach is novel because it accommodates short distances between layers via continuous diffusion coefficient updates.

The remainder of this paper is structured as follows. Section \ref{Sec:SystemModel} presents the system model and problem formulation. Section \ref{results} offers results and discussions of our findings. Finally, Section \ref{sec:conclusion} concludes the paper.

\section{System Model}
\label{Sec:SystemModel} 
\begin{figure}[!t]
	\centering
	\includegraphics[width=2.3in]{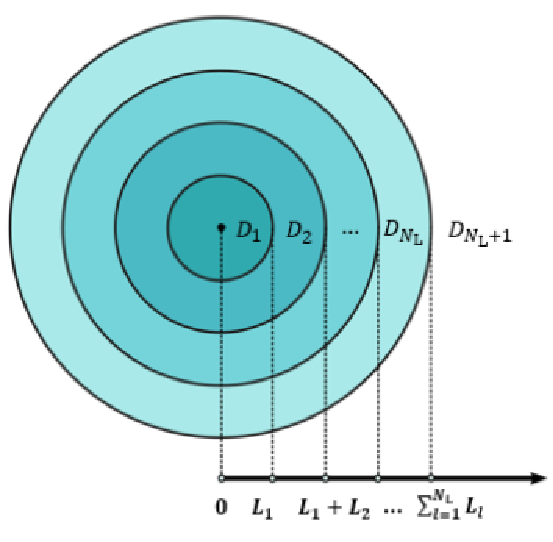}
    	\caption{Cross-section of a multi-layer spherical structure with $N_{\mathrm{L}}$ layers. Each layer $i$ is characterized by a diffusion coefficient $D_i$ and width $L_i$.} 
	\label{centrekd}
\end{figure}

Our system model has a non-homogeneous porous multicellular sphere consisting of \( N_{\mathrm{L}} \) finite layers each with width $L_i$, $i \in \{1, 2, 3,..., N_{\mathrm{L}}\}$, enclosed within an additional infinite layer (i.e., the space outside the sphere is unbounded). This yields a system with a total of \( N_{\mathrm{L}} + 1 \) layers, as shown in Fig. \ref{centrekd}. 
The \textit{porosity parameter} of each layer, $\varepsilon_i$, serves as the ratio of the extracellular space of the corresponding layer to its overall volume, i.e.,  
$\varepsilon_i=1-\frac{N_{\mathrm{c},i}V_{\mathrm{c},i}}{V_{\mathrm{L},i}}$, where $N_{\mathrm{c},i}$, $V_{\mathrm{c},i}$, and $V_{\mathrm{L},i}$ are the number of cells within layer $i$, volume of each cell, and volume of layer $i$, respectively.
We assume that the sphere is immersed in an unbounded fluid medium with zero flow rate (extendable to a system with convection). that fills its extracellular space. 
The effective diffusion within the whole porous sphere volume is reduced compared to the free fluid diffusion outside \cite{Rezaei2024}. Each layer has its own molecule degradation rate $k_i$ and effective molecule diffusion coefficient $D_{\textrm{eff},i}$, simply $D_i$ for the rest of the paper, according to its porosity parameter $\varepsilon_i$.
$D_i$ can be determined from the molecules' free fluid diffusion coefficient $D$, i.e., $D_i=\frac{\varepsilon_i}{\tau_i}D$, where $\tau_i$ is the tortuosity of layer $i$ and it refers to the degree of path irregularity or curvature experienced by a molecule while it traverses through the extracellular space of each layer of the spheroid \cite{FRENNING201188}. $\tau_i$ is a function of porosity, i.e., $\tau_i=\frac{1}{\left(\varepsilon_i\right)^{0.5}}$.

To describe the environment geometry, we use the spherical coordinate system. The system origin is positioned at the center of the multi-layered structure, with $\bar r=(r,\theta,\varphi)$ representing radial, elevation, and azimuthal coordinates, respectively.
At the interface between two diffusive environments that have different diffusion coefficients, a non-homogeneous continuity condition for flow must be satisfied, which is expressed as 
\begin{equation}\label{BC2}
D_i \frac{\partial c_i(\bar r,t)}{\partial r} =D_{i+1} \frac{\partial c_{i+1}(\bar r,t)}{\partial r}, ~~~i \in\{1, 2, ..., N_{\mathrm L}\},
\end{equation}
and another non-homogeneous boundary condition for a fully permeable interface (i.e., molecules crossing an interface are assumed to pass through without reflecting off of the interface) that is generally modeled as \cite[Ch. 3]{crank1979mathematics}
\begin{equation}\label{BC1}
c_i(\bar r,t)=\kappa_ic_{i+1}(\bar r,t), ~~~~~~~~~~~~~~~~i \in\{1, 2, ..., N_{\mathrm L}\},
\end{equation}
must also be satisfied, where $ \bar r \in \partial \Omega$, $\partial \Omega$ denotes the sphere boundary region, $\Omega$ is the spheroid region, and $c_i$ is the concentration function inside layer $i$. 
The constant $\kappa_i$ is found as $\sqrt{\frac{D_{i+1}}{D_i}}$, for $i \in\{1:N_{\mathrm L}+1\}$ \cite{arjmandi1}.
Thus, for $\kappa_i\neq 1$, a concentration discontinuity (i.e., jump) occurs at the boundary.

An impulse source (transmitter) may be located within any of these layers. For the $p$th layer containing a Dirac delta type impulsive source at location $\bar{r}=\bar{r}_0$, the improper Green's function in the frequency domain is denoted by \( G^p(\bar{r}|\bar{r}_0;\omega) \), where \( p \in \{1, 2, \ldots, N_{\mathrm{L}}+1\} \) and $\omega$ is the frequency. Improper Green functions, also known as generalized Green functions, are used when the impulsive source is inside the layer or on its boundaries. These functions facilitate the analysis of complex geometries, such as layered solids, by satisfying in-homogeneous boundary conditions at some or all interfaces. In contrast, proper Green functions are employed when the source is outside the region and must satisfy homogeneous boundary conditions, meaning they are zero at the boundaries. In layers without the source, we instead use the channel impulse response \( U^q(\bar{r},\bar{r}_0;\omega) \), where \( q \neq p \), with a non-reciprocal argument, since this is neither a proper nor an improper Green function \cite{mandelis2013diffusion}. 
The isotropic configuration yields the following radial equations for different regions. For a region \( i \) containing the source, the governing equation including the source time-modulation factor $e^{\boldsymbol{i}\omega t_0}$ in the definition is: 
\begin{equation}\label{fickG}
D_i{\nabla^2}G^i(r|r_0;\omega;t_0) - \sigma_i^2(\omega) G^i(r|r_0;\omega;t_0) = -S(r|r_0;\omega;t_0),
\end{equation} 
where \( i \in \{1, 2, \dots, N_{\mathrm{L}} + 1\} \) and $S(r|r_0;\omega;t_0) = \frac{\delta(r - r_0)\delta(\theta-\theta_0)\delta(\varphi-\varphi_0)e^{-\boldsymbol{i}\omega t_0}}{r^2\sin\theta} \, \si{m^{-3}}$ defines the impulse source at (coordinates) releasing molecules at time $t=t_0$, and $\boldsymbol{i}=\sqrt{-1}$,  and $\sigma_i=\sqrt{\frac{{k_i +\boldsymbol{i}\omega}}{D_i}}$. The solution \( G^i \) in this region represents the isotropic Green's function. For any region \( j \), where the source is not located (\( j \neq i \)), the equation becomes:
\begin{equation}\label{fickU}
D_j{\nabla^2}U^j(\bar{r}, \bar{r}_0 ;\omega; t_0) - \sigma_j^2(\omega) U^j(\bar{r}, \bar{r}_0 ;\omega; t_0)) = 0,
\end{equation}
where $\sigma_{\rm j}=\sqrt{\frac{k_j+\boldsymbol{i}\omega}{D_j}}$. The general solutions for the Green's function in region $i$ with the impulsive source and impulse response in region $j$  ($j \neq i$) are given as follows, respectively:
\begin{equation}\label{EqsG}
\begin{aligned}
G^i(r, \theta, \varphi| \bar{r}_0; \omega )
& = \sum\limits_{n = 0}^\infty \sum\limits_{m = 0}^n H_{mn} g_n^i(r, \omega) \cos \big(m(\varphi - \varphi_0)\big) \\
& \quad \times P_n^m(\cos \theta), 
\end{aligned}
\end{equation}

and 
\begin{equation}\label{EqsU}
\begin{aligned}
U^j(r, \theta, \varphi, \bar{r}_0 ;\omega)
& = \sum\limits_{n = 0}^\infty \sum\limits_{m = 0}^n H_{mn} u_n^j(r, \omega) \cos \big(m(\varphi - \varphi_0)\big) \\
& \quad \times P_n^m(\cos \theta). 
\end{aligned}
\end{equation}

In \((\ref{EqsG})\) and \((\ref{EqsU})\), \( g_n^i(r, \omega) \) and \( u_n^j(r, \omega) \) denote the unknown radial Green's function for layer \( i \) and the unknown impulse response for layer \( j \), respectively. These functions must be determined by solving the corresponding equations and their boundary conditions. In these equations, $\mathcal{F}(\theta,\varphi)=\cos(m \varphi)P_n^m(\cos\theta)$ is a spherical harmonic with degree $n$ and order $m$ that  satisfies the  partial differential equation (PDE)
\begin{equation}\label{logender22}
\begin{aligned}
&\frac{D}{{\sin \theta }}\frac{\partial }{{\partial \theta }}\left(\sin \theta \frac{{\partial \mathcal{F}(\theta,\varphi)}}{{\partial \theta }}\right) + \frac{D}{{{{\sin }^2}\theta }}\frac{{{\partial ^2}\mathcal{F}(\theta,\varphi)}}{{\partial {\varphi ^2}}}\\
&+{n(n+1)}\mathcal{F}(\theta,\varphi)=0,
\end{aligned}
\end{equation}
and $H_{mn}$ denotes the unknown coefficient that needs to be determined for the particular configuration. 
Also, the representation of
${\delta (\varphi  - {\varphi _0})} \frac{{\delta (\theta  - {\theta _0})}}{{\sin \theta }}$  based on the aforementioned spherical harmonic is given by  
\begin{equation}\label{deltaphitheta1}
\begin{aligned}
&{\delta (\varphi  - {\varphi _0})} \frac{{\delta (\theta  - {\theta _0})}}{{\sin \theta }} =\sum\limits_{n=0}^\infty  \sum\limits_{m = 0}^n \lambda_m \frac{{2n + 1}}{2}\frac{{(n - m)!}}{{(n + m)!}}\\
&\times {P_n^m(\cos \theta )P_n^m(\cos {\theta _0})} {\cos \big(m(\varphi  - \varphi _0)\big)},
\end{aligned}
\end{equation}
where $\lambda_0=\frac{1}{2\pi}$ and $\lambda_m=\frac{1}{\pi}, m\geq 1$.
\begin{figure*}
	\small
	\begin{equation}\label{matchg}
	\begin{array}{l}
	\sum\limits_{n=0}^{\infty}\sum\limits_{m=0}^{n} H_{mn} \frac{1}{{{r^2}}}\frac{\partial }{{\partial r}}\left({r^2}\frac{{\partial g_n^i(r,\omega)}}{{\partial r}}\right) P_n^m(\cos\theta)\cos\left(m(\varphi-\varphi_{0})\right) +\sum\limits_{n=0}^{\infty}\sum\limits_{m=0}^{n} H_{mn} g_n^i(r,\omega) \frac{1}{{{r^2}\sin \theta }}\frac{\partial }{{\partial \theta }}\left(\sin \theta \frac{{\partial P_n^m(\cos\theta)\cos\left(m(\varphi-\varphi_{0})\right)}}{{\partial \theta }}\right) +\\ \frac{1}{{{r^2}{{\sin }^2}\theta }}\frac{{{\partial ^2}\cos\left(m(\varphi-\varphi_{0})\right)}P_n^m(\cos\theta)}{{\partial {\varphi ^2}}}
	-\left(\sigma_i^2({\boldsymbol{i}\omega})\right)\sum\limits_{n=0}^{\infty}\sum\limits_{m=0}^{n} H_{mn}g_n^i(r,\omega) P_n^m(\cos\theta)\cos\left(m(\varphi-\varphi_{0})\right)\\
	= 	-\frac{1}{D_i}\sum\limits_{n=0}^\infty  \sum\limits_{m = 0}^n 
	\lambda_m \frac{{2n + 1}}{2}\frac{{(n - m)!}}{{(n + m)!}}
	\times{P_n^m(\cos \theta )P_n^m(\cos {\theta _{0}})} {\cos \left(m(\varphi  - \varphi _{0})\right)}\delta(r-r_0).
	\end{array}
	\end{equation}
	\color{black}
	\hrulefill
\end{figure*}
\begin{figure*}
\small
	\begin{equation}\label{matchu}
	\begin{array}{l}
\sum\limits_{n=0}^{\infty}\sum\limits_{m=0}^{n} H_{mn} \frac{1}{{{r^2}}}\frac{\partial }{{\partial r}}\left({r^2}\frac{{\partial u_n^j(r,\omega)}}{{\partial r}}\right) P_n^m(\cos\theta)\cos\left(m(\varphi-\varphi_{0})\right) +\sum\limits_{n=0}^{\infty}\sum\limits_{m=0}^{n} H_{mn} u_n^j(r,\omega) \frac{1}{{{r^2}\sin \theta }}\frac{\partial }{{\partial \theta }}(\sin \theta \frac{{\partial P_n^m(\cos\theta)\cos\left(m(\varphi-\varphi_{0})\right)}}{{\partial \theta }}) +\\ \frac{u_n^j}{{{r^2}{{\sin }^2}\theta }}\frac{{{\partial ^2}\cos(m(\varphi-\varphi_{0}))P_n^m(\cos\theta)}}{{\partial {\varphi ^2}}}
	-(\sigma_j^2(\boldsymbol{i}\omega))\sum\limits_{n=0}^{\infty}\sum\limits_{m=0}^{n} H_{mn}u_n^j(r,\omega) P_n^m(\cos\theta)\cos\left(m(\varphi-\varphi_{0})\right)= 	0.
	\end{array}
	\end{equation}
	\color{black}
	\hrulefill
\end{figure*}
Replacing $G^i$, $U^j$ and ${\delta (\varphi  - {\varphi _0})} \frac{{\delta (\theta  - {\theta _0})}}{{\sin \theta }}$  in \eqref{fickG} and \eqref{fickU} by the corresponding series-form representations given by \eqref{EqsG}, \eqref{EqsU} and \eqref{deltaphitheta1}, respectively, leads to \eqref{matchg} and \eqref{matchu} at the top of the following page. Matching the two sides of \eqref{matchg} yields
\begin{equation}
H_{mn}=\lambda_m \frac{{2n + 1}}{2}\frac{{(n - m)!}}{{(n + m)!}}P_n^m(\cos\theta_0),
\end{equation} 
and
\begin{equation}\label{RPDEg}
\begin{aligned}
&r^2\frac{\partial^2 g_n^i(r,\omega)}{\partial r^2}+2r\frac{\partial g_n^i(r,\omega)}{\partial r}\\
&+((\sigma^{i})^2r^2-n(n+1))g_n^i(r,\omega)=\frac{1}{D_i}\delta ( r -r_0).
\end{aligned}
\end{equation} 
Similarly,  \eqref{matchu} is reduced to
\begin{equation}\label{RPDEu}
\begin{aligned}
&r^2\frac{\partial^2 u_n^j(r,\omega)}{\partial r^2}+2r\frac{\partial u_n^j(r,\omega)}{\partial r}\\
&+((\sigma^{j})^2r^2-n(n+1))u_n^j(r,\omega)=0.
\end{aligned}
\end{equation}
The solutions of the homogeneous forms of the PDE of \eqref{RPDEg} and \eqref{RPDEu} are then given by
\begin{align}
g_n^{i}(r,\omega) &= \begin{cases}
    A_n^i J_n(kr) + B_n^i Y_n(kr), & R_{i-1} < r < r_0 \\
    C_n^i J_n(kr) + E_n^i Y_n(kr), & r_0 < r < R_i
\end{cases} \label{Rr01} \\
u_n^{j}(r,\omega) &= A_n^j J_n(kr) + B_n^j Y_n(kr), \quad R_{j-1} < r < R_j \label{Rr02}
\end{align}
where $R_{l^{\prime}} = \sum_{l=1}^{l^{\prime}} L_l$ for $l^{\prime} \geq 0$, with $R_0 = 0$ and the width of the $({N_{L}+1})$th layer is $\infty$. 
In these two equations, $J_n(\cdot)$ and $Y_n(\cdot)$ are the $n$th order of the first and second kinds of spherical Bessel function, respectively. Solution \eqref{Rr01} applies to a layer with a source at \( r_0 \), covering the sublayers \( R_{i-1} < r < r_0 \) and \( r_0 < r < R_i \), whereas solution \eqref{Rr02} applies to each layer $j$ without a source in the region \( R_{j-1} < r < R_j \).
It should be noted that the coefficient \( B_n^i \) of \( Y_n(kr) \) is set to zero in the interior solution (i.e., layer or sub-layer near the center) to prevent the field from ``blowing up'' at the origin. This is because \( Y_n(kr) \) becomes unbounded as \( r \to 0 \). Also, for the outer layer (or sublayer) that is bounded at infinity, it is preferable to use the solution \( A_n h_n(kr) \), where \( h_n(kr) \) is the spherical Hankel function of the third kind and of order \( n \) \cite{mandelis2013diffusion}. The coefficients $A_n^i$, $B_n^i$, $C_n^j$, $E_n^j$, $A_n^j$, and $B_n^j$ are determined using the following boundary conditions:
\begin{align}
D_i\frac{\partial T^i(r,\omega)}{\partial r }\bigg|_{ r=R_i^-} &= D_{i+1}\frac{\partial T^{i+1}(r,\omega)}{\partial r }\bigg|_{r =R_i^+} \label{B1}, \\
T^i(r,\omega)\bigg|_{ r=R_i^-} &= \kappa_{i} T^{i+1}(r,\omega)\bigg|_{ r=R_i^+}, \label{B2}
\end{align}
where $T^x=g_n^x$ if the source is located in layer $x$ with width $L_x$, and $T^x=u_n^x$ if the source is not located in layer $x$. The subscripts $i$ and $i + 1$ indicate that the two corresponding layers are adjacent (i.e., coupled), and superscripts + and - represent a minuscule distance above or below the corresponding interface, respectively. Additionally, at \( r = r_0 \), the continuity of the Green's function \( g_n \) and the discontinuity of its derivative must be satisfied, as described in \cite{mandelis2013diffusion}:
\begin{equation}\label{SD}
r^2 \frac{\partial g_n^i(r,\omega)}{\partial r}\bigg|_{r=r^{+}_{0}}
- r^2 \frac{\partial g_n^i(r,\omega)}{\partial r}\bigg|_{r=r^{-}_{0}}=\frac{1}{D_i},
\end{equation}
and 
\begin{equation}\label{SC}
g_n^i(r,\omega)\bigg|_{r=r^{+}_{0}}
= g_n^i(r,\omega)\bigg|_{r=r^{-}_{0}}.
\end{equation} 
\subsection{Special Case: Three-Layer Spherical Structure in an Infinite Medium}
While the general solution for $N_{\mathrm{L}}$ layers has been presented in the previous section, it is instructive to examine the specific case of a three-layer structure embedded in an infinite medium. This configuration, comprising four distinct regions (three finite layers and one infinite outer layer), occurs frequently in practical applications such as tumors and deserves careful attention.
We denote the regions as $\{1,2,3,4\}$, where region 4 refers to the outer infinite medium. The concentration distributions in these regions are governed by two equations:
\begin{itemize}    
\item Eq.~\eqref{fickG} applies to the layer containing the impulse source   \item Eq.~\eqref{fickU} describes the layers without a source
\end{itemize}
Similarly, the boundary conditions and interface continuity between each pair of adjacent layers are given by \eqref{BC2} and \eqref{BC1}.
Consider the impulse source located in layer 1. The general solution for the Green's function in region 1 is then given by \eqref{EqsG}. For any other layer without a source ($j \neq 1$), the general solution for the impulse response is given by \eqref{EqsU}.
Based on the discussion for Eqs.~\eqref{Rr01} and \eqref{Rr02}, the homogeneous part of the PDEs, $g_n^1$ in region 1 in \eqref{EqsG}, and $u_n^j$ for layers $j \in \{2,3,4\}$ without a source in ~\eqref{EqsU}, are given by the following equations, respectively:
\begin{equation} 
g_n^{1}(r,\omega) = \begin{cases} A_n^1J_n(kr), & 0 < r < r_0 \\ C_n^1J_n(kr) + E_n^1 Y_n(kr), & r_0 < r < R_1    
\end{cases}
\end{equation}
\begin{align} u_n^{2}(r,\omega) &= A_n^2J_n(kr)+B_n^2 Y_n(kr), & R_{1}<r<R_2 \\
 u_n^{3}(r,\omega) &= A_n^3J_n(kr)+B_n^3 Y_n(kr), & R_{2}<r<R_3 \\ u_n^{4}(r,\omega) &= A_n^4h_n(kr). & r>R_{3}
\end{align}
The parameters $A_n^1$, $C_n^1$, $E_n^1$, $A_n^2$, $B_n^2$, $A_n^3$, $B_n^3$, and $A_n^4$ are determined subject to the following boundary and continuity conditions:
\begin{align}
D_{1}\frac{\partial g_n^1(r,\omega)}{\partial r }\bigg|_{ r=R_1^-} &= D_{2}\frac{\partial u_n^{2}(r,\omega)}{\partial r }\bigg|_{r =R_1^+} \label{B1_1}, \\
g_n^1(r,\omega)\bigg|_{ r=R_1^-} &= \kappa_{1}\times u_n^{2}(r,\omega)\bigg|_{ r=R_1^+}, \label{B2_1}
\end{align}
\begin{align}
D_{2}\frac{\partial u_n^2(r,\omega)}{\partial r }\bigg|_{ r=R_2^-} &= D_{3}\frac{\partial u_n^{3}(r,\omega)}{\partial r }\bigg|_{r =R_2^+} \label{B1_2}, \\
u_n^2(r,\omega)\bigg|_{ r=R_2^-} &= \kappa_{2}\times u_n^{3}(r,\omega)\bigg|_{ r=R_2^+}, \label{B2_2}
\end{align}
\begin{align}
D_{3}\frac{\partial u_n^3(r,\omega)}{\partial r }\bigg|_{ r=R_3^-} &= D_{4}\frac{\partial u_n^{4}(r,\omega)}{\partial r }\bigg|_{r =R_3^+} \label{B1_3}, \\
u_n^3(r,\omega)\bigg|_{ r=R_3^-} &= \kappa_{3}\times u_n^{4}(r,\omega)\bigg|_{ r=R_3^+}, \label{B2_3}
\end{align}
\begin{equation}\label{SD_exp}
r^2 \frac{\partial g_n^1(r,\omega)}{\partial r}\bigg|_{r=r^{+}_{0}}
- r^2 \frac{\partial g_n^1(r,\omega)}{\partial r}\bigg|_{r=r^{-}_{0}}=\frac{1}{D_1},
\end{equation}
\begin{equation}\label{SC_exp}
g_n^1(r,\omega)\bigg|_{r=r^{+}_{0}}
= g_n^1(r,\omega)\bigg|_{r=r^{-}_{0}}.
\end{equation}

\section{Simulation and Numerical Results}
\label{results}
This section presents the outcomes derived from the analytical model detailed in Section \ref{Sec:SystemModel}. It is important to emphasize that our model is designed for application to a wide range of spherical multi-layer structures. However, for demonstration and validation, we apply the model to a large, non-homogeneous spheroid with distinct layers of varying porosity. Our model incorporates the average of radially-dependent porosity values for each layer from \cite{goodman2008spatio} to account for this heterogeneity. To clarify the presentation, we focus on a three-layer spheroid with radius $275\,\mu\rm m$, where the porosities of the layers of the same thickness are $\varepsilon_1 = 0.2964$, $\varepsilon_2 = 0.1196$, and $\varepsilon_3 = 0.1697$, respectively. However, the model can be used for any multi-layered structure with an arbitrary number of layers. 
Also, the diffusion coefficient for molecules in the free fluid is considered to be $D=10^{-9}\,\si{cm^2/s}$. We first verify our proposed analysis for diffusion problems in multi-layered spherical structures with PBS. Then,  we investigate how the porosity of one layer affects the molecule concentration in other layers. 
The PBS is implemented in MATLAB (R2021b; The MathWorks, Natick, MA) where the time is divided into time steps of $\Delta t=0.5\,\si{s}$. 
The molecules released from the transmitter move independently in the 3-D space. The displacement of a molecule in $\Delta t$ is modeled as a Gaussian random variable with zero mean and variance $2D\Delta t$ in each Cartesian dimension. 

For the PBS, each layer $i$ of the spheroid is simply a diffusion environment with the effective coefficient $D_i$. Whenever a molecule trajectory intersects with an interface between any two layers, we adjust the molecule's displacement vector after the intersection point according to the corresponding diffusion coefficient. When the displacement vector calculated for a molecule intersects the boundary between two interfaces in a given time slot, the length of the portion of the vector inside the second layer is updated based on the effective diffusion coefficient $D_{i}^{\textrm{new}}$, i.e., that part of the vector is scaled by the factor $\sqrt{\frac{D_{i}^{\textrm{new}}}{D_{i}^{\textrm{old}}}}$. 
We also account for instances of molecules crossing multiple boundaries within one time interval, where we adjust a molecule's displacement vector for every crossing according to the corresponding diffusion coefficient.

\begin{figure}[!t]
	\centering
	\includegraphics[width=3.8in]{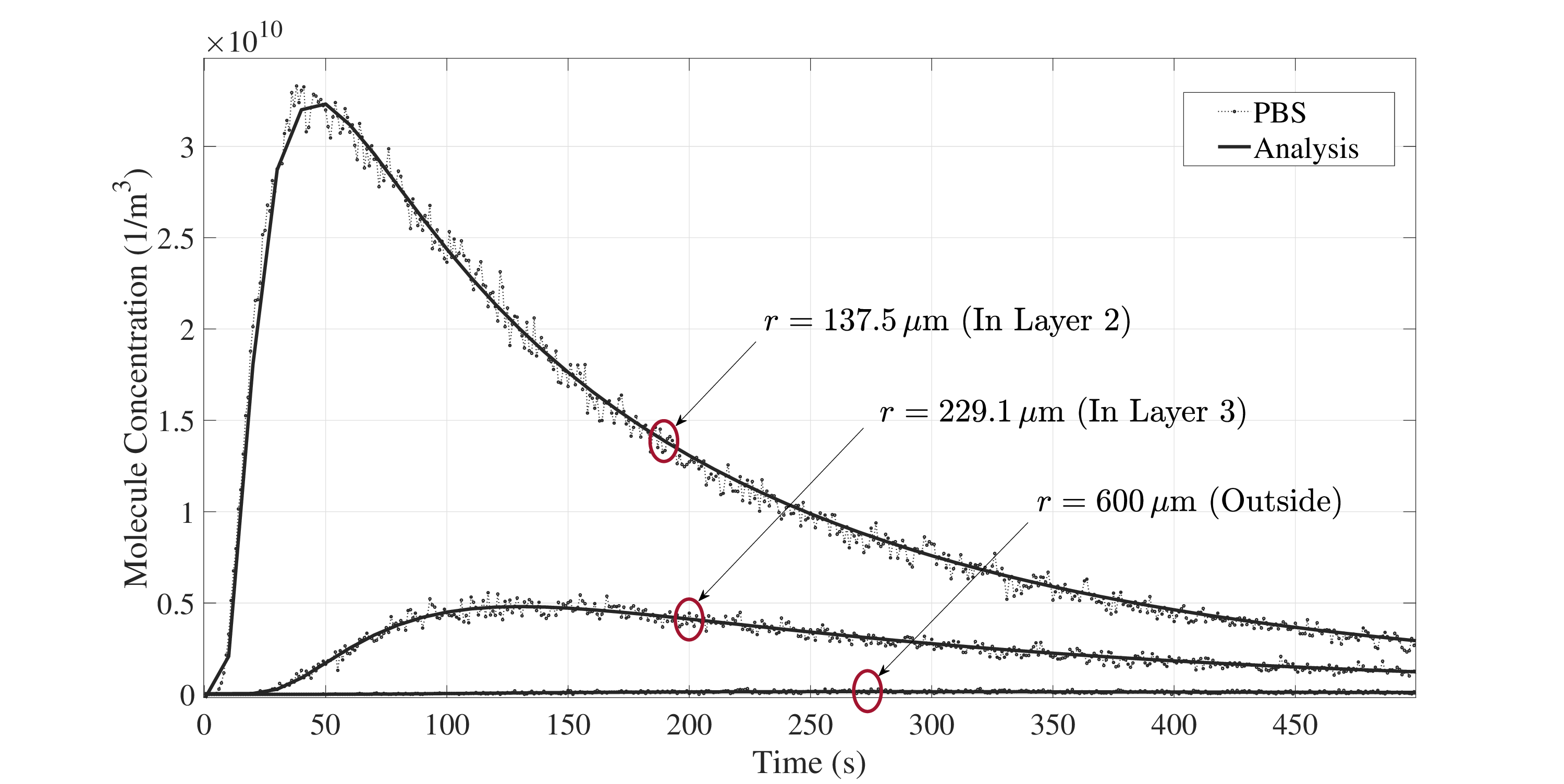}
	\caption{Concentration profiles at points $\bar{r}=(r\, \mathrm{\mu m}, \pi/2, 0)$ across two distinct spheroid layers and the external medium, where $r$ values are specified in the figure. A point transmitter is located at $\bar{r}_0 = (45.83\, \mathrm{\mu m}, \pi/2, \pi/2)$.}
	\label{ConctxL1}
\end{figure}
A molecule may also be absorbed by cells from the spheroid's extracellular space during a time step $\Delta t$ with approximate probability $(1 - \exp(-k_i\cdot \Delta t))$, where $k_i$ is the molecule degradation rate in layer $i$. To isolate the effect of porosity on molecule propagation, we set $k_i = 0$ for all layers in our analysis. To measure a point concentration at a given time and location, we place a transparent sphere centered at that location with a small radius of $10\,\si{\mu m}$ and count the number of molecules inside the sphere at that time. The concentration would then be the counted number of molecules divided by the volume of the sphere. Then the concentration value is normalized, i.e., divided by the number of molecules.
\begin{figure*}[t]
    \centering
    \begin{minipage}[t]{.48\textwidth}
        \includegraphics[width=\linewidth]{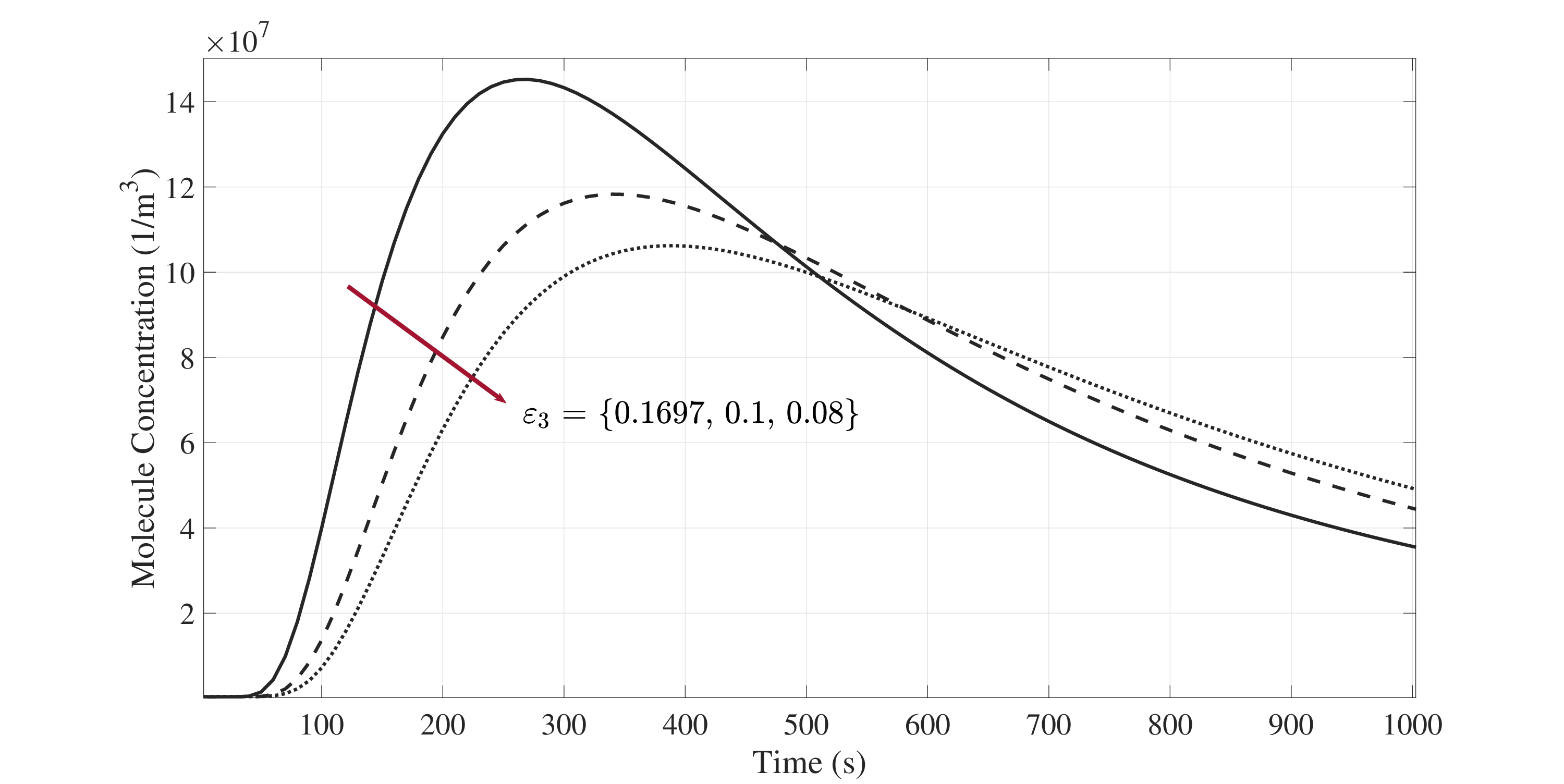}
        \caption{Impact of the outer layer (layer 3) porosity on molecular diffusion outside a three-layer spheroid, with the transmitter positioned at $\bar{r}_0 = (45.83\, \mathrm{\mu m}, \pi/2, \pi/2)$.}
        \label{Fig_Rporosity}
        \includegraphics[width=\linewidth]{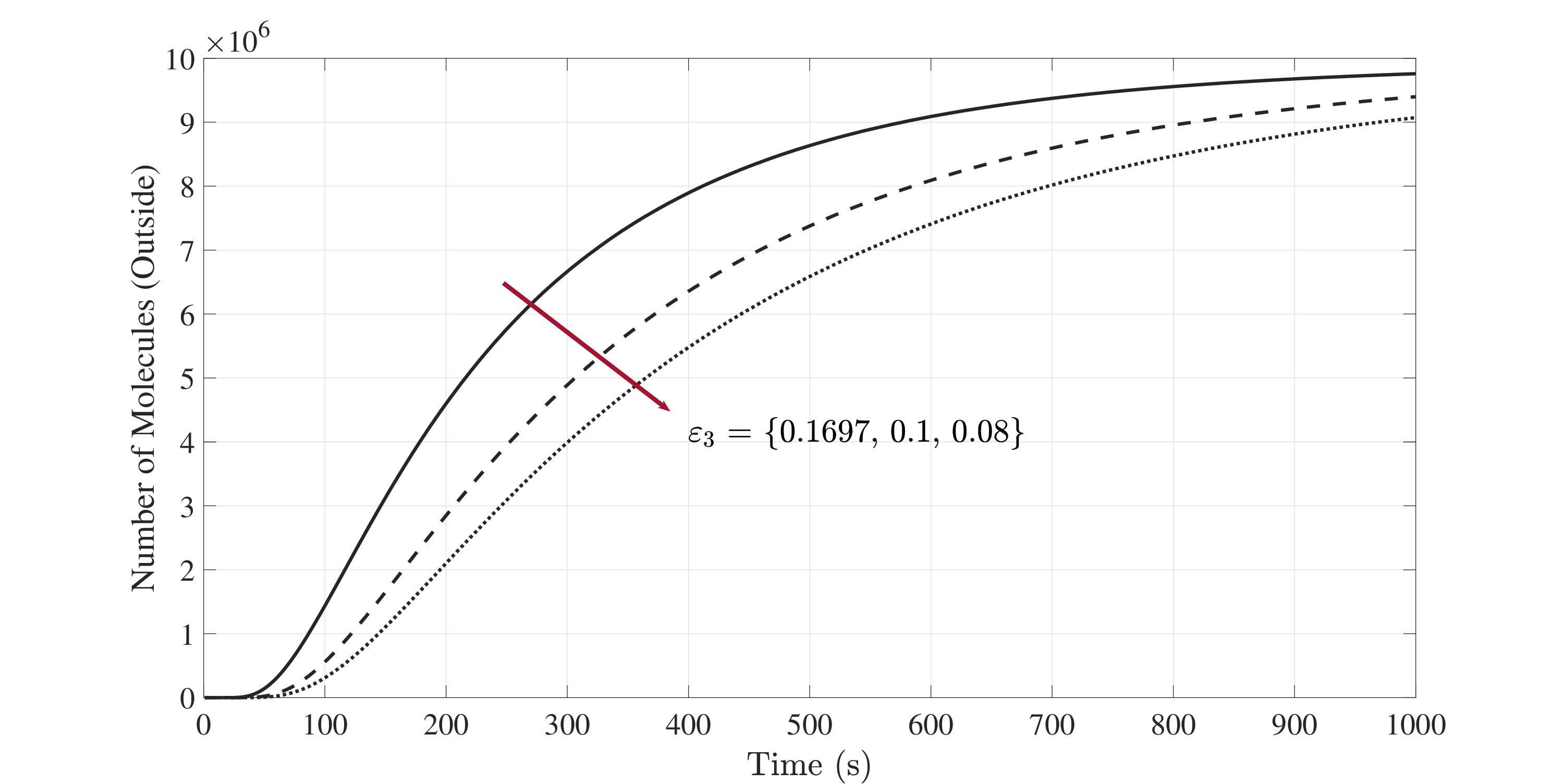}
        \caption{Temporal evolution of total molecule count outside the spheroid for three different layer 3 porosity values, simulated using PBS, with the transmitter located at $\bar{r}_0 = (45.83\, \mathrm{\mu m}, \pi/2, \pi/2)$.}
        \label{3eps.total}
    \end{minipage}\hfill
    \begin{minipage}[t]{.48\textwidth}
        \includegraphics[width=\linewidth]{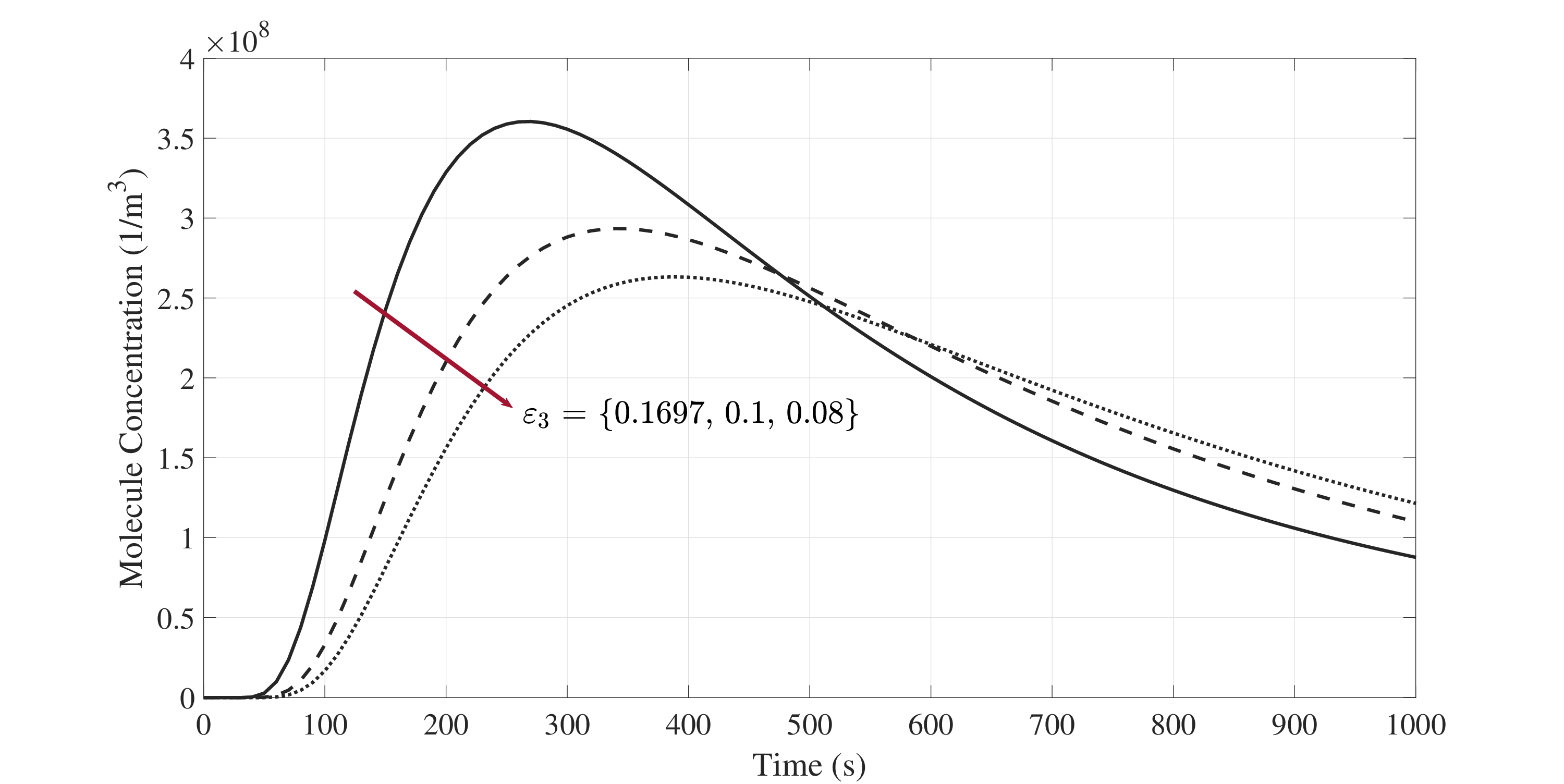}
        \caption{Impact of the outer layer (layer 3) porosity on molecular diffusion within the inner layer (layer 1) of a three-layer spheroid, with the transmitter positioned at $\bar{r}_0 = (600\, \mathrm{\mu m}, \pi/2, \pi/2)$.}
        \label{Fig_Rporosity2}
        \includegraphics[width=\linewidth]{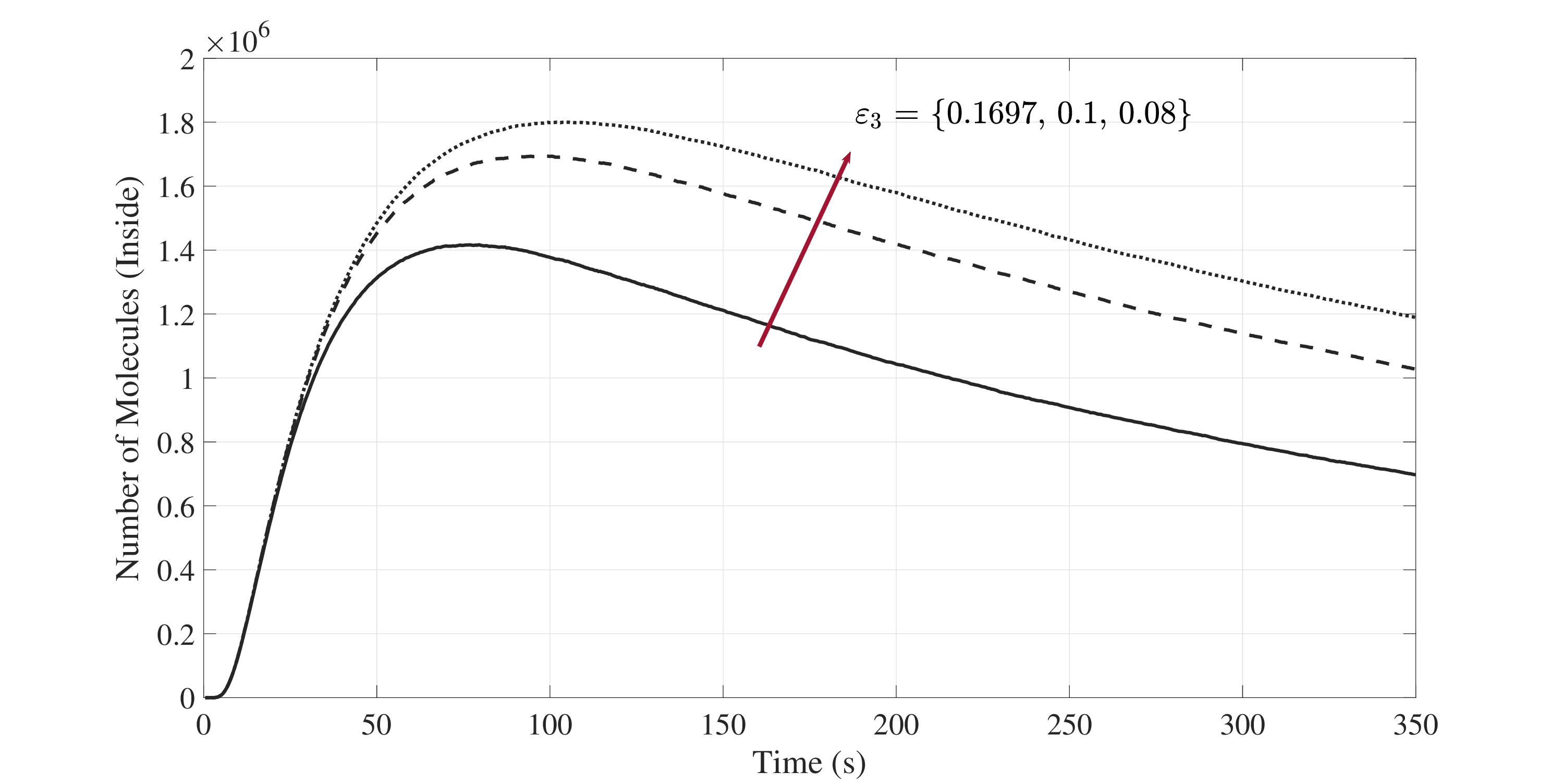}
        \caption{Temporal evolution of molecular count within the spheroid for three different porosity values of layer 3, simulated using PBS, with the transmitter located at $\bar{r}_0 = (600\, \mathrm{\mu m}, \pi/2, \pi/2)$.}
        \label{3eps.total2}
    \end{minipage}
\end{figure*}

Fig. \ref{ConctxL1} illustrates the concentration profiles at different points across spheroid layers and in the external medium, given an instantaneous point transmitter located at $\bar{r}_0=(45.83\, \si{\mu m}, \pi/2, \pi/2)$, derived from the analytical model presented in Section \ref{Sec:SystemModel} and PBS. As observed over a sufficiently long measurement period, the analytical results are fully supported by the PBS. Near the point source transmitter, the concentration profile peak is highest and decreases with distance, and its peak approaches zero at the periphery of the spheroid due to the low porosity of the spheroid. These findings have particular relevance in drug delivery applications, particularly for implant-based systems within tumors. The observed concentration gradient proposes high efficacy in the target area near the transmitter, as well as reduced toxicity in surrounding tissues due to minimal drug presence at the periphery. 

Fig. \ref{Fig_Rporosity} shows the influence of reduced porosity in the outer layers of a multi-layered spheroid on molecular diffusion in the surrounding medium, where the transmitter is located at $\bar{r}_0=(45.83\, \si{\mu m}, \pi/2, \pi/2)$. A decrease in the outer layer's porosity, particularly in layer 3, results in delayed signal arrival, lower peak concentrations, and increased signal dispersion outside the spheroid. These effects are mostly due to the enhanced barrier properties of the less porous outer layer. As porosity decreases, molecules face greater difficulty traversing these layers, which leads to delayed and attenuated peak signals. Furthermore, the reduced porosity hinders molecular escape from the spheroid, which contributes to increased signal dispersion in the surrounding medium. Fig. \ref{3eps.total} shows the temporal growth of total molecule count outside the spheroid for three distinct layer 3 porosity values, as simulated using PBS. Lower porosity in layer 3 significantly delays the outward molecule penetration, even without considering interior degradation rates. This delay results in prolonged molecule retention within the spheroid layers.

Fig. \ref{Fig_Rporosity2} demonstrates how reduced porosity in the outer layer (e.g., layer 3) of the multi-layer spherical model affects molecular diffusion to inner layers (e.g., layer 1). In this scenario, the transmitter is positioned outside the spheroid at $\bar{r}_0=(600\, \si{\mu m}, \pi/2, \pi/2)$. Lower outer porosity results in delayed signals, lower peaks, and increased dispersion in inner layers. This is due to the increased difficulty for molecules to both enter and exit the spheroid through less porous outer layers. In Fig. \ref{3eps.total2}, we examine the impact of layer 3 porosity on the total number of molecules within the spheroid when they are released from an external transmitter. Interestingly, lower porosity in layer 3 results in a higher molecule count inside the spheroid. This occurs because while reduced porosity hinders molecules from entering inner layers, it also traps them within layer 3, impeding their escape back into the surrounding medium. Consequently, decreasing porosity in this case also leads to increased molecular retention within the spheroid, primarily concentrated in the outer layer.

\section{Conclusion}
\label{sec:conclusion}

We developed a mathematical formulation for a diffusive molecular communication (MC) system in multi-layered spherical channels. The model accommodates any number of layers and allows for arbitrary positioning of transmitters and receivers. We derived analytical solutions for the Green's function in the layer with a point source and the channel impulse response in layers without sources. As confirmed by PBS, we investigated how different porosity values of one layer can impact the diffusion within other layers. Low porosity in outer layers restricts outward diffusion from inner sources and inward diffusion from external sources. These results underscore the critical role of outer layer porosity in molecular transport dynamics both within and around multi-layered spheroids.
This is particularly significant for optimizing release rates in drug delivery systems and designing advanced drug carriers to minimize toxicity to surrounding tissues. Future research will focus on extending the framework to other non-parallel geometries. Additionally, we aim to investigate drug release mechanisms based on layer-specific physiological parameters, such as pH levels in cancer spheroids.

\ifCLASSOPTIONcaptionsoff
  \newpage
\fi
\bibliographystyle{IEEEtran}
\bibliography{REFICC1}

\begin{thebibliography}{10}
\providecommand{\url}[1]{#1}
\csname url@samestyle\endcsname
\providecommand{\newblock}{\relax}
\providecommand{\bibinfo}[2]{#2}
\providecommand{\BIBentrySTDinterwordspacing}{\spaceskip=0pt\relax}
\providecommand{\BIBentryALTinterwordstretchfactor}{4}
\providecommand{\BIBentryALTinterwordspacing}{\spaceskip=\fontdimen2\font plus
\BIBentryALTinterwordstretchfactor\fontdimen3\font minus \fontdimen4\font\relax}
\providecommand{\BIBforeignlanguage}[2]{{%
\expandafter\ifx\csname l@#1\endcsname\relax
\typeout{** WARNING: IEEEtran.bst: No hyphenation pattern has been}%
\typeout{** loaded for the language `#1'. Using the pattern for}%
\typeout{** the default language instead.}%
\else
\language=\csname l@#1\endcsname
\fi
#2}}
\providecommand{\BIBdecl}{\relax}
\BIBdecl

\bibitem{8742793}
V.~Jamali, A.~Ahmadzadeh, W.~Wicke, A.~Noel, and R.~Schober, ``Channel modeling for diffusive molecular communication—a tutorial review,'' \emph{Proc. of the IEEE}, vol. 107, no.~7, pp. 1256--1301, July 2019.

\bibitem{mustam2017multilayer}
S.~M. Mustam, S.~K. Syed~Yusof, and S.~Nejatian, ``Multilayer diffusion-based molecular communication,'' \emph{Trans. on Emerging Telecommunications Technologies}, vol.~28, no.~1, p. e2935, 2017.

\bibitem{Composite}
M.~M. Al-Zu’bi and A.~M. Sanagavarapu, ``Modeling a composite molecular communication channel,'' \emph{IEEE Trans. Commun.}, vol.~66, no.~8, pp. 3420--3433, 2018.

\bibitem{carr2016semi}
E.~J. Carr and I.~W. Turner, ``A semi-analytical solution for multilayer diffusion in a composite medium consisting of a large number of layers,'' \emph{Appl. Math. Model.}, vol.~40, no. 15-16, pp. 7034--7050, 2016.

\bibitem{rodrigo2016solution}
M.~R. Rodrigo and A.~L. Worthy, ``Solution of multilayer diffusion problems via the {L}aplace transform,'' \emph{J. Math. Anal. Appl.}, vol. 444, no.~1, pp. 475--502, 2016.

\bibitem{pontrelli2013local}
G.~Pontrelli, A.~Di~Mascio, and F.~De~Monte, ``Local mass non-equilibrium dynamics in multi-layered porous media: application to the drug-eluting stent,'' \emph{Int. J. Heat Mass Transf.}, vol.~66, pp. 844--854, 2013.

\bibitem{jain2022theoretical}
A.~Jain, S.~McGinty, G.~Pontrelli, and L.~Zhou, ``Theoretical modeling of endovascular drug delivery into a multilayer arterial wall from a drug-coated balloon,'' \emph{Int. J. Heat Mass Transf.}, vol. 187, p. 122572, 2022.

\bibitem{carr2018modelling}
E.~J. Carr and G.~Pontrelli, ``Modelling mass diffusion for a multi-layer sphere immersed in a semi-infinite medium: application to drug delivery,'' \emph{Math. Biosci.}, vol. 303, pp. 1--9, 2018.

\bibitem{kaoui2018mechanistic}
B.~Kaoui, M.~Lauricella, and G.~Pontrelli, ``Mechanistic modelling of drug release from multi-layer capsules,'' \emph{Comput. Biol. Med.}, vol.~93, pp. 149--157, 2018.

\bibitem{mahesh2022mathematical}
N.~Mahesh, N.~Singh, and P.~Talukdar, ``A mathematical model for understanding nanoparticle biodistribution after intratumoral injection in cancer tumors,'' \emph{J. of Drug Delivery Science and Technology}, vol.~68, p. 103048, 2022.

\bibitem{al2020modelling}
M.~Al-Zu’bi and A.~Mohan, ``Modelling of combination therapy using implantable anticancer drug delivery with thermal ablation in solid tumor,'' \emph{Sci. Rep.}, vol.~10, no.~1, p. 19366, 2020.

\bibitem{nath2016three}
S.~Nath and G.~R. Devi, ``Three-dimensional culture systems in cancer research: Focus on tumor spheroid model,'' \emph{Pharmacol. Ther.}, vol. 163, pp. 94--108, 2016.

\bibitem{Rezaei2024}
M.~Rezaei, H.~Arjmandi, M.~Zoofaghari, K.~Kanebratt, L.~Vilén, D.~Janzén, P.~Gennemark, and A.~Noel, ``Spheroidal molecular communication via diffusion: Signaling between homogeneous cell aggregates,'' \emph{IEEE Trans. Mol. Biol. Multi-Scale Commun.}, vol.~10, no.~1, pp. 197--210, 2024.

\bibitem{FRENNING201188}
G.~Frenning, ``Modelling drug release from inert matrix systems: From moving-boundary to continuous-field descriptions,'' \emph{Int. J. Pharm.}, vol. 418, no.~1, pp. 88--99, 2011, mathematical modeling of drug delivery systems:.

\bibitem{crank1979mathematics}
J.~Crank, \emph{The mathematics of diffusion}.\hskip 1em plus 0.5em minus 0.4em\relax Oxford University Press, 1979.

\bibitem{arjmandi1}
\BIBentryALTinterwordspacing
H.~Arjmandi, K.~P. Kanebratt, L.~Vilén, P.~Gennemark, and A.~Noel, ``3d cell aggregates amplify diffusion signals,'' \emph{PLOS ONE}, vol.~19, no.~9, pp. 1--18, 09 2024. [Online]. Available: \url{https://doi.org/10.1371/journal.pone.0310109}
\BIBentrySTDinterwordspacing

\bibitem{mandelis2013diffusion}
A.~Mandelis, \emph{Diffusion-wave fields: mathematical methods and Green functions}.\hskip 1em plus 0.5em minus 0.4em\relax Springer Science \& Business Media, 2013.

\bibitem{goodman2008spatio}
T.~T. Goodman, J.~Chen, K.~Matveev, and S.~H. Pun, ``Spatio-temporal modeling of nanoparticle delivery to multicellular tumor spheroids,'' \emph{Biotechnol. Bioeng.}, vol. 101, no.~2, pp. 388--399, 2008.

\end{thebibliography}
\end{document}